\documentclass[fleqn,10pt]{wlscirep}
\usepackage[utf8]{inputenc}
\usepackage[T1]{fontenc}
\usepackage{lineno}

\usepackage{amsmath}
\usepackage{bm}
\usepackage{ulem}
\usepackage{soul}
\usepackage{comment}

\title{Ultra-low threshold lasing through phase front engineering via a metallic circular aperture}

\author[1,*]{Zhixin Wang}
\author[1]{Filippos Kapsalidis}
\author[1]{Ruijun Wang}
\author[1]{Mattias Beck}
\author[1,*]{J\'er\^ome Faist}
\affil[1]{ETH Z\"urich, Institute of Quantum Electronics, Auguste-Piccard-Hof 1, Z\"urich 8093, Switzerland}

\affil[*]{corresponding author: Zhixin Wang (zhixwang@phys.ethz.ch); J\'er\^ome Faist (jfaist@ethz.ch)}

\begin{abstract}

Semiconductor lasers with ultra-low thresholds and minimal footprints are a topic of active research \cite{mccall1992whispering,painter1999two,hill2007lasing, gu2017semiconductor, crosnier2017hybrid,ma2019applications}. Such devices require a combination of high quality factor laser cavities with small active region volumes, which drives the quest for novel cavity geometries exploiting nano-optic concepts \cite{akahane2003high,oulton2009plasmon,khajavikhan2012thresholdless,lu2012plasmonic,kress2017customizable}. For high-reflectivity coated ridge lasers, where light is tightly confined in the waveguide, a low threshold can only be achieved by strongly reducing the diffraction losses arising at the laser facet. We show here that, somewhat counter-intuitively, opening a carefully designed aperture in a metallic facet coating can simultaneously enhance both its transmission and modal reflectivity by correcting the phase front at the subwavelength scale. Numerical simulations and experimental results demonstrate a reduction of optical mirror loss by up to 40\% while the transmission is increased by four orders of magnitude. Applying this approach to both facets of a short cavity quantum cascade laser, we achieve laser operation at room temperature with an electrical dissipation of only 143~mW. Such light sources are especially suitable for portable and battery-operated chemical agent sensing applications operating in the mid-infrared wavelength range, where multiple greenhouse and pollutant gases have their fundamental absorption lines. Our work suggests possibilities for further applications including frequency comb dispersion engineering, and can be implemented in a broad range of optoelectronic systems.

\end{abstract}

\begin{document}

\flushbottom
\maketitle

\thispagestyle{empty}

\section*{Introduction}
Progress in fabrication technologies has allowed the engineering of optical fields at subwavelength scales \cite{yu2008small,novotny2012principles, chen2020flat}, leading to the discovery of many novel phenomena such as extraordinary optical transmission, in which an above-unity normalized-to-area transmittance is achieved by introducing a two-dimensional (2D) periodic array of subwavelength holes into a metallic film \cite{ebbesen1998extraordinary}. A good understanding of these phenomena requires going beyond the traditional Kirchhoff paradigms \cite{born2013principles}. As proposed by Bethe in 1944, the diffraction field of a small metallic aperture can be considered as excited by magnetic and electric moments \cite{bethe1944theory}. Substantial progress in both theoretical and experimental aspects has been made since then \cite{bouwkamp1950diffraction, cohn1952microwave,roberts1987electromagnetic,astilean2000light,takakura2001optical,yang2002resonant,de2002light,nikitin2008electromagnetic,adam2008advanced,garcia2010light}, but mainly focused on transmission enhancement. In this work, we demonstrate that subwavelength apertures also provide a way to engineer the reflectivity of the metal film by tailoring the retarded phases of the beam. 

Many mid-infrared optical chemical sensors can operate with optical powers much below a milliwatt \cite{palaferri2018room, sterczewski2020mid}. Given the fact that intersubband and interband cascade lasers have demonstrated wall-plug efficiencies well above 10 \% \cite{kim2015high,wang2020room}, lasers for such portable sensors would demand only tens of milliwatt electrical power to operate, which simply requires the development of suitable laser cavities. 

\begin{figure}[hpt!]
\centering\includegraphics[width=\textwidth]{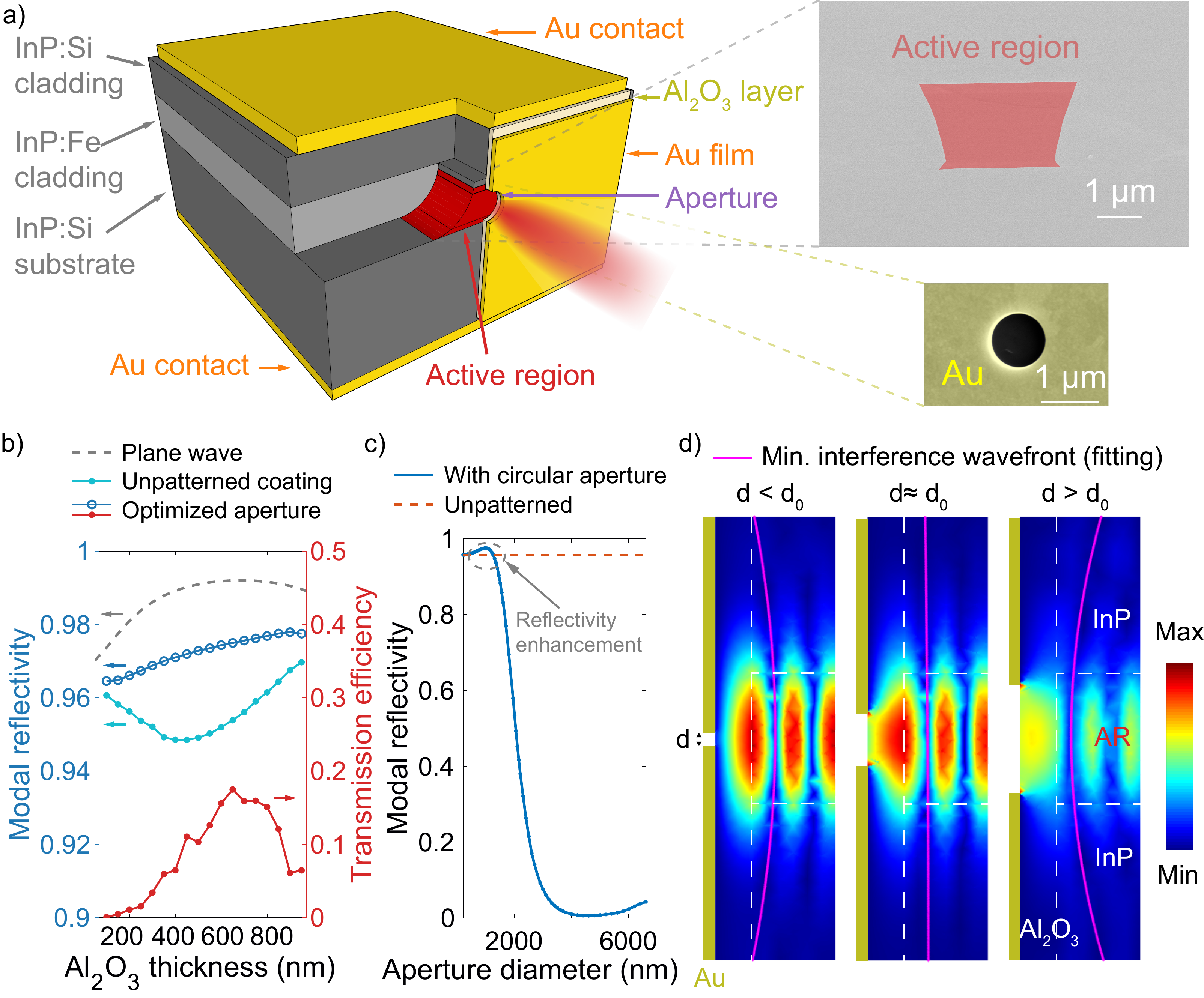}
\caption{(a) A schematic drawing of the laser with an aperture in the front-side metallic coating. Here only half of the front coating is shown and the active region is drawn as exposed for a clear illustration. In the actual device, the active region is fully buried by the cladding and the coating fully covers the facet. The upper right inset shows an SEM image of the laser facet before deposition of the coating. The lower right inset shows the SEM image of the circular aperture in the metallic coating, with a diameter of around 950 nm, created by the focused ion beam milling. (b) Modal reflectivity of the metallic coating as a function of the insulating layer thickness, for three scenarios, including the ideal plane-wave case (grey, dashed line), the case with a laser waveguide but without an aperture (light blue dotted line), and the case with a waveguide and a circular aperture through the metallic film of which the aperture size is optimized for each Al$_2$O$_3$ thickness (dark blue circled line). The red dotted line shows the transmission efficiency as a function of the Al$_2$O$_3$ thickness for the case with the circular aperture. The wavelength of light is 4.5 $\mu$m. (c) Modal reflectivity of the metallic coating as a function of the circular aperture diameter, where the insulating layer thickness is set as $\lambda/4$. The straight dashed red line shows the reflectivity of the coating without an aperture. The dashed grey circle shows the region where the modal reflectivity is enhanced by the presence of the aperture.  (d) Side-view plots of the electric field  ($abs(E_z)$) in a simulation with three aperture sizes. The pink lines are the fitted curves for the minimum interference, which indicates the shape of the wave-front. Parameter $d_0$ is the aperture diameter with the maximum modal reflectivity. The aperture diameters in these three simulations are 250 nm, 1000 nm and 2000 nm, respectively. }
\label{fig:refl_sim}
\end{figure}

Because of the unfavorable polarization selection rules for intersubband transitions, a vertical cavity surface emitting laser \cite{wilmsen2001vertical}, which is a usual choice for small volume and low dissipation devices, is not a geometry that can be easily implemented for intersubband quantum cascade lasers (QCLs) \cite{faist2013quantum}. A promising alternative structure, as shown in Fig. \ref{fig:refl_sim}(a), is a short, tightly-confining narrow-ridge active region, in our case a QCL processed into a buried heterostructure with ridges down to 1 $\mu$m width. In this configuration, keeping the optical losses to an acceptable value requires facet coatings  ideally approaching > 98\% modal reflectivity. 

The large value of the dielectric constant of noble metals in the mid-infrared makes them a very attractive solution to achieve highly reflective coatings, especially if they are combined with a low dielectric constant insulator deposited on the semiconductor. Indeed, as shown by the grey dashed line in Fig. \ref{fig:refl_sim}(b) where the reflectivity of the coating is reported as a function of dielectric thickness, for a plane wave at normal incidence with the wavelength of 4.5 $\mu$m, a maximum reflectivity of 99.2\% can be achieved by a quarter-wave thick layer of Al$_2$O$_3$ (700 nm) followed by a 200 nm thick Au metalization. The light-blue solid-dotted line on the same graph is the reflectivity computed considering the real geometry of the waveguide and the confined mode profile. In fact, when the diffraction losses at the semiconductor-dielectric interface are taken into account, the computed reflectivity drops to 95.6\%. In addition, this coating cannot be used as an outcoupler since the transmission is on the order of 10$^{-7}$. 

Paradoxically, opening a subwavelength aperture on the facet can increase both transmission and reflection, enabling the coating to be used simultaneously as an outcoupler and a high reflector. Indeed, as shown in Fig. \ref{fig:refl_sim}(b) by the dark-blue circled line, opening a circular aperture with an optimized diameter increases the reflectivity of the facet. Values of the aperture diameters at each dielectric thickness are shown in Supplemental Material Sec. A. As an example, for an Al$_2$O$_3$ thickness of 700 nm, an aperture with 950 nm diameter increases the reflectivity to 97.5\%, reducing the optical losses by over 40\%. 

The transmission efficiency of the patterned coating, defined as the transmitted power divided by non-reflected power [$T/(1-R)$], is reported as the red dotted line in Fig. \ref{fig:refl_sim}(b). 
At the Al$_2$O$_3$ thickness of 700 nm, for example, the transmission efficiency is 15.9\% and the corresponding transmissivity is 0.4\%, four orders of magnitude higher than the one of the unpatterned case.

For a fixed $\lambda/4$ (700 nm) Al$_2$O$_3$ thickness, the modal reflectivity as a function of the aperture diameter is shown in Fig. \ref{fig:refl_sim}(c). The reflectivity enhancement phenomenon is clearly observed in the subwavelength region, as marked by the dashed grey circle. In contrast, a minimum in reflectivity, which is close to zero, demonstrates the transmission resonance phenomenon as the aperture diameter becomes comparable to the laser wavelength \cite{ebbesen1998extraordinary,de2002light, garcia2010light}. 

The reflectivity enhancement by the aperture can be understood from the perspective of phase compensation: the retarded nature of the excited magnetic and electric moments around the subwavelength-sized metallic aperture add a phase that corrects for the natural divergence of the beam.  Figure \ref{fig:refl_sim}(d) shows the simulated side-view image of the field patterns ($E_z$) with three different aperture sizes. The incident and the reflected waves from the waveguide and the coating interfere, and the curvature of the local minima (dark regions) indicates the shape of the wave-front, as shown by the fitted pink curves in the figures. For small aperture sizes, the amplitude of the back-scattered field is nearly constant and the modal reflectivity is maximized when the wavefront becomes flat. As shown in Fig. \ref{fig:refl_sim}(d), this situation arises when the aperture diameter $d \approx d_0 = 1$ $\mu$m. In this case, the flat wavefront indicates that the diffraction phase delay is significantly compensated by the aperture resonance. As also shown in Fig. \ref{fig:refl_sim}(d) , when the aperture size is smaller or larger than the optimal dimension, the interference wave-front exhibits a positive or negative curvature. Further simulations on the detail of the field patterns are shown in the  Supplemental Material Sec. B.

\section*{Experimental results and discussion}
Shown in the upper right inset of Fig. \ref{fig:refl_sim}(a) is a scanning electron microscope (SEM) image of the bare facet of one of our narrow-ridge buried heterostructure devices, to which we then apply a high-reflectivity coating (see Methods section). Using a procedure also described in the Methods section, a circular aperture with 950~nm diameter is opened using a focused ion beam milling instrument. As shown by the SEM image shown in Fig.  \ref{fig:refl_sim}(a), the aperture displays a very clean geometry with smooth edges. Alignment between the aperture and the waveguide has been achieved either using a set of etched markers or energy-dispersive X-ray spectroscopy. See Supplementary Material Sec. C for more details.

\begin{figure}[hpt!]
\centering\includegraphics[width=0.8\textwidth]{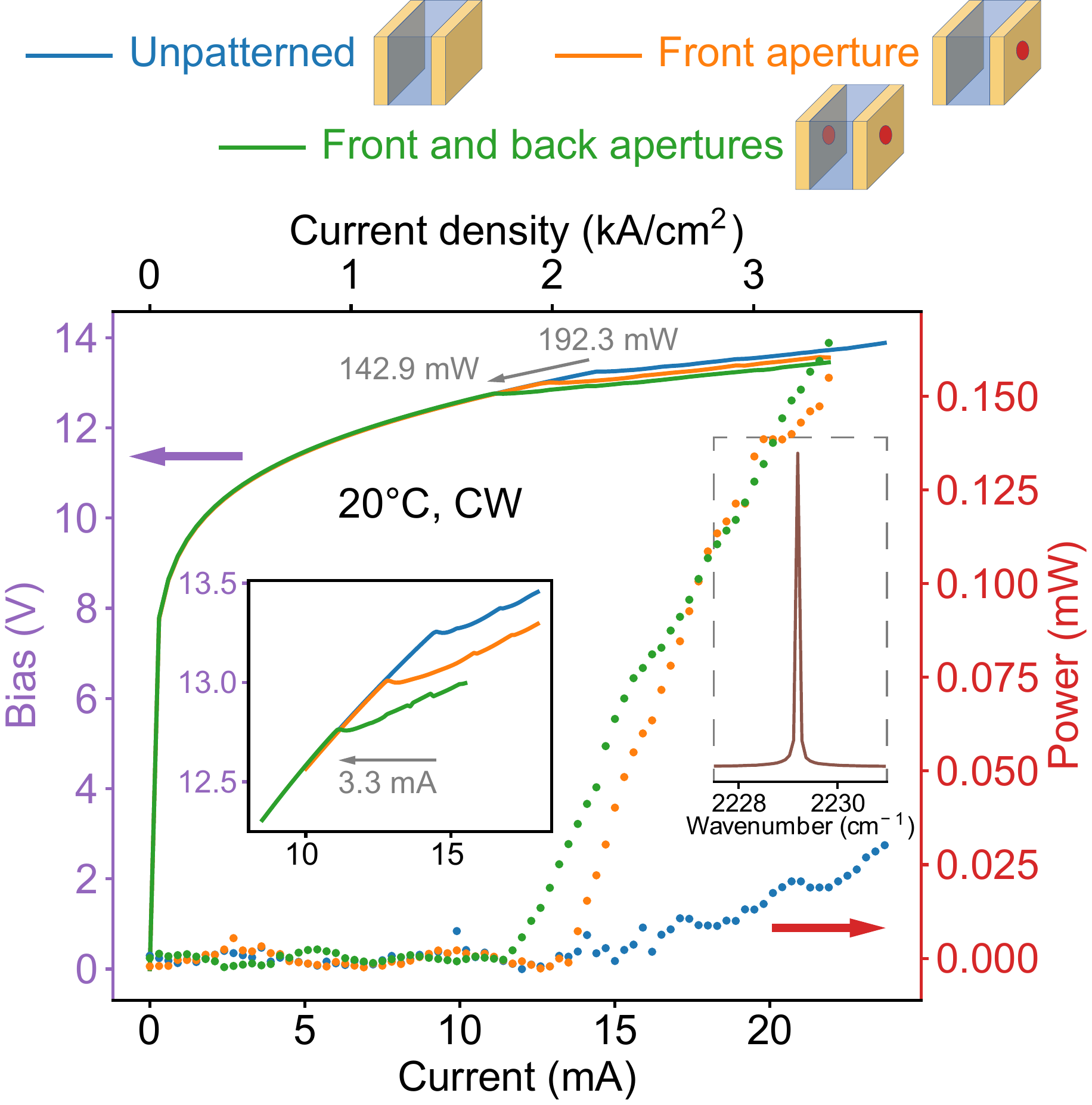}
\caption{Continuous-wave current-voltage-power characterization of the short QCL with metallic coatings on both facets, measured at room temperature. The blue line and dots show the results where no apertures are patterned on the coatings. The orange line and dots show the results where a circular aperture with a diameter of 950 nm is milled through the Au film of the front coating. The green line and dots show the results where another circular aperture with the diameter of 950 nm is patterned in the backside high-reflectivity coating. The current step of the measurements is 0.3 mA. The reduction of the threshold current is shown by the movement of the current-voltage kinks towards the left, and are more clearly displayed in the left inset, where the measurements are conducted using a finer current step of 0.1 mA. In total, the threshold dissipation power is reduced by over 25\%. The laser is single-mode at threshold, as shown in the right inset, where the spectral resolution of the measurement is 0.075 cm$^{-1}$.}
\label{fig:IthReduce}
\end{figure}

To experimentally confirm the results of these simulations, a relatively short QCL cleaved to a length of 265 $\mu$m is coated on both facets using an Al$_2$O$_3$/Au coating with the thicknesses of Al$_2$O$_3$ and Au being around 700 nm and 200 nm, respectively. See Supplementary Material Sec. C for the SEM image of the whole device.  
Shown in Fig. \ref{fig:IthReduce} (blue dotted and solid lines) are the light output and applied bias as a function of the injected current in continuous-wave operation for a device with a 2.5 $\mu$m-wide ridge and unpatterned coatings. A  threshold at 14.5 mA is clearly indicated by the kink in the current-voltage curve (blue solid line). Since the amplitude of the output power is smaller than the thermal drift of the measurement setup, the lasing threshold cannot be identified from the current-power curve (blue dots).

After this first characterization, a circular aperture with a diameter of around 950 nm is then milled through the Au layer of the front coating, and is aligned with the laser waveguide as shown in the bottom right inset of Fig. \ref{fig:refl_sim}(a). The voltage and power characteristics as a function of injected current after the milling are reported by the orange solid line and dots in Fig. \ref{fig:IthReduce}. 
Strikingly, and as predicted by our simulations, the lasing threshold is now reduced by 11 \%, down to 12.9 mA, commensurate with the predicted 40\% decrease in front-facet mirror loss. In addition, an optical power of 150 $\mu$W is now extracted from the front facet as the transmission of the coating is increased from 10$^{-7}$ to 10$^{-3}$. Consequently, while still more accurately indicated by the kink in the current-voltage curve, the threshold is now clearly visible in the current-power curve (orange dots).

By duplicating the same aperture-milling procedure in the backside coating, the lasing threshold is further reduced to 11.2 mA, as shown by the green solid line and dots in Fig. \ref{fig:IthReduce}. The left inset of Fig. \ref{fig:IthReduce} shows the current-voltage characteristics with a finer measurement step of 0.1 mA, and a total reduction of 3.3 mA in the threshold current is clearly observed. Compared to the initial threshold dissipation with unpatterned coatings, the final threshold dissipation is decreased by 25 \%, down to 143 mW at 20 $^{\circ}$C, which is 45 \% lower than the previously achieved record \cite{cheng2020ultralow}. At -20$^{\circ}$C, the threshold current is 8.3 mA. Furthermore, we emphasize that the presented threshold-reducing technique is reproducible. See Supplemental Material Sec. D for the data measured 
with additional temperatures and devices.

\begin{figure}[hpt!]
\centering\includegraphics[width=\textwidth]{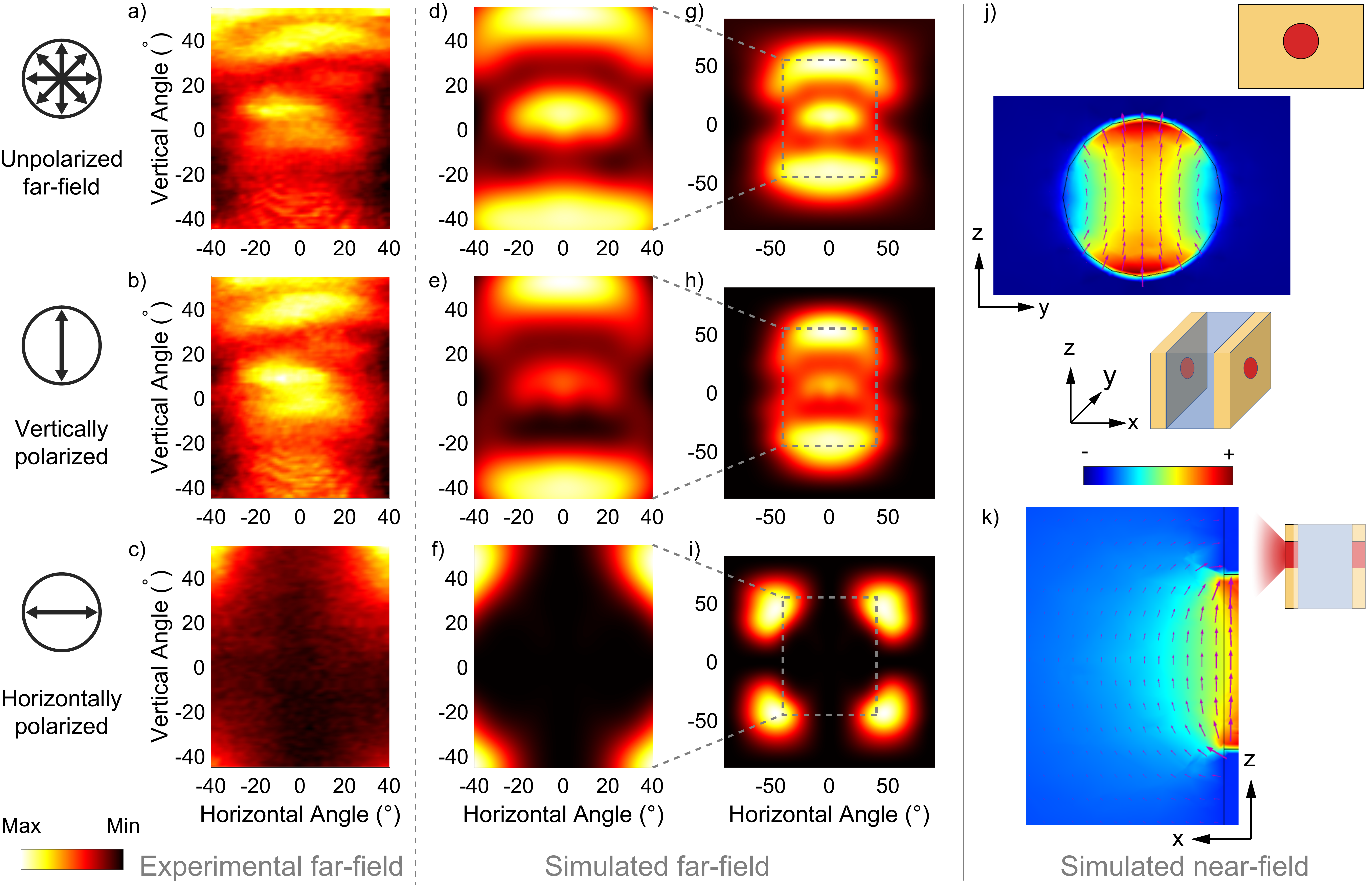}
\caption{(a-i): Measured (a-c) and simulated (d-i) far-field patterns of the beam emitted through the circular aperture in the front metallic coating. (a,d,g) are unpolarized far-field beam patterns, (b,e,h) are vertically polarized and (c,f,i) are horizontally polarized. (a-c) are measured at room temperature with a bias voltage of 13.6 V. Plotted in (a-i) are the patterns of the energy ($|E|^2$). (j,k): Simulated near-field patterns near the circular aperture in the front view (j) and the side view (k). The colormap shows the vertical component of the electric field ($E_z$) and the vectors represent the electric field vectors. Upper right insets indicate the direction of the view in each plot. The plane of (j) is selected as the outer interface of the metallic coating and the plane of (k) is the cut-plane through the middle of the device. }
\label{fig:ffp}
\end{figure}

The measured far-field patterns and the polarization characteristics of lasing beams are reported in Figs. \ref{fig:ffp}(a-c), which agree with the simulated results  shown in Figs. \ref{fig:ffp}(d-i). The beam is mainly polarized in the vertical direction, agreeing with the selection rule of intersubband transitions \cite{faist2013quantum}. Despite  the circular symmetry of the aperture, the far-field pattern is highly asymmetrical and is much broader in the vertical direction. As shown by Figs. \ref{fig:ffp}(j,k) which illustrate the near-field patterns, the electric field inside the aperture is oriented similarly as the one of a subwavelength dipole antenna \cite{balanis2016antenna}. As a result of the large diffraction angle, the vector Stratton-Chu model \cite{stratton1939diffraction, stratton2007electromagnetic} must be applied for the far-field calculation and the scalar Kirchhoff theory \cite{goodman2005introduction,yariv2007photonics,born2013principles} is no longer valid. As a result, the beam divergence in the vertical direction is much larger than the one along the horizontal direction. The presence of the metallic aperture strongly enhances the relative intensity of the far field in the horizontal polarization created by the field discontinuity at the edges of the waveguide, as shown with more details in the Supplemental Material Sec. E.

\section*{Conclusion}
In a rather counter-intuitive way, we demonstrated that engineering the near field at a laser facet by opening a metallic aperture enabled a very significant reduction of the optical cavity losses of a semiconductor laser and therefore a strong reduction of its electrical dissipation. Quite obviously, the devices presented here as an example are not fully optimized. For example, using facets realized via dry etching techniques would enable much shorter cavities to be fabricated. Furthermore, by tuning the size and the shape of the aperture, this work could be further extended to applications such as frequency comb dispersion compensation and far-field engineering.

\section*{Methods}
\subsection*{Details on the fabrication process}
In this work, the active region of the QCL is grown by molecular beam epitaxy (MBE) and is comprised of a strained In$_{0.66}$Ga$_{0.34}$As/Al$_{0.665}$In$_{0.335}$As heterogeneous quantum cascade stack of two active regions, one centered at 2325 cm$^{-1}$ (4.3 $\mu$m) and the other at 2174 cm$^{-1}$ (4.6 $\mu$m), published previously in Ref. \cite{suess2016dual}. The QCL is fabricated into a buried heterostructure configuration \cite{beck2002continuous,suess2016advanced}.  A distributed-feedback (DFB) grating is etched on top of the active region, with the grating period spread around a range of target wavelengths. The active region is patterned in ridges of 1-3 $\mu$m by wet etching. The side (InP:Fe) and top (InP:Si) claddings surrounding the active region are grown by metal-organic vapour phase epitaxy (MOVPE) regrowth. After patterning the top and bottom Ohmic contacts, the devices are mechanically cleaved (length $\approx$ 250 $\mu$m) and mounted either epitaxial side up on copper mounts, or epitaxial side down on aluminum nitride substrates. The metallic coatings, consisting of Al$_2$O$_3$ and Au layers, are grown by an electron beam evaporation process.

\subsection*{Details on the focused ion beam milling}
The focused ion beam (FIB) milling technique is conducted using the Helios 5 UX (Ga ion) manufactured by ThermoFisher and maintained by ScopeM, ETH Zurich. For each facet, the FIB milling is used in two steps: before the deposition of the coating, alignment markers are milled by FIB on the laser facet about  20 $\mu$m away from the laser waveguide.  After the coating deposition, the circular aperture is milled through the metallic layer with the help of the markers. The milling stops at the interface between the Al$_2$O$_3$ and Au layers. In the end, the alignment between the aperture and the waveguide is checked by a high-resolution scanning electron microscope (SEM) or energy-dispersive X-ray spectroscopy (EDX).

Images of the markers and the EDX are shown in the Supplemental Material, Sec. C.

\subsection*{Measurement devices}
For all experimental measurements, the laser is mounted on a Laboratory Laser Housing from Alpes Lasers. For the current-voltage-power characterization and the spectrum measurement, the laser is driven by a Keithley 2420 source-meter {in  continuous-wave operation}. The current and voltage are measured by the same source-meter. The spectrum is measured with a Fourier-transform infrared (FTIR) spectrometer  (Bruker, Vertex 80). For the far-field measurement, the laser is driven by an Agilent 8114A pulse generator. The far-field patterns are measured using a pyroelectric detector (Gentec-EO: THZ2I-BL-BNC) mounted on a motorized scanning stage. A DSP lock-in amplifier (Model 7265 by EG\&G Instruments) is also used for the far-field measurement.

\bibliography{MyCitation}







\section*{Correspondence} 
*Correspondence should be addressed to Z. Wang (email: zhixwang@phys.ethz.ch) and J\'er\^ome Faist (email: jfaist@ethz.ch).

\end{document}